# Comparison of N-body Simulations to Statistical Observations of Galaxy Pairs

Roger E. Bartlett[1] and Jane C. Charlton[1,2]

## ABSTRACT

N-body simulations were conducted of pairs of galaxies with a 3:1 mass ratio on parabolic orbits, in order to quantify the effect of dynamical friction. The effects of varying the ratio of the dark matter halo size to the distance of closest approach were explored. Once the dark matter halos are fully overlapping, the more massive simulated galaxies achieve a larger maximum separation after the first encounter despite the increased dynamical friction caused by the more extended halos. Projected separation and radial velocity histograms were generated by "observing" the simulation results at various times and from various orientations. These histograms were compared with observations of galaxy pairs (Charlton & Salpeter 1991; Chengalur, Salpeter, & Terzian 1993) with the result that large halo radii ($\sim 200-600$kpc) and wide distances of closest approach are generally favored. It is difficult to reconcile the small radial velocity differences that have been observed (median of $\sim 30$ km/s, Chengalur et al. 1993) with the simulations when we sample all parts of the orbits equally. Including an additional population of wide pairs that have just recently reached "turnabout" from the Hubble flow would lower the median velocity differences. Models suggest that additional data for pairs at intermediate separations should have a somewhat larger median velocity difference than the wide pairs. Very narrow pairs include galaxies which are interacting, and whose gaseous components respond to forces other than gravity. If consistently small $\Delta v$ are measured from neutral hydrogen velocities in a larger sample of narrow pairs, pressure forces and dissipation effects on the gaseous components could be responsible.

*Subject headings:* dark matter — galaxies: clustering

---

[1]Astronomy and Astrophysics Department, Pennsylvania State University, University Park, PA 16802

[2]Center for Gravitational Physics and Geometry, Pennsylvania State University

## 1. Introduction

The present distribution of galaxy pairs is a synthesis of pairs at various stages of orbits that are a consequence of unknown initial conditions. The evolution of a particular pair will involve galaxies (or proto-galaxies) that initially separate due to Hubble flow, reach turn-about and head towards first approach, overshoot merger, and continue to bounce back and forth for an indeterminate number of times. The details of this process are influenced by the initial orbital parameters and by the effect of dynamical friction due to overlap of dark matter halos.

Charlton and Salpeter (1991) (hereafter CS) analyzed the CfA (Davis et al. 1982; Davis & Huchra 1982; Huchra et al. 1983) and SSRS (da Costa et al. 1988) redshift survey catalogs, focusing on pairs in low density regions. It was found that pairs with radial velocity differences, $\Delta v$, less than 150km/s are quite likely to be bound, and that bound pairs exist out to separations as large as 1Mpc. The distribution of projected separations was approximately flat from 0 to 1Mpc, and this was interpreted as evidence for a merger/replenishment scenario for pairs at small separations. This picture implies that the dark matter halos of galaxies in low density regions must be large enough (a few hundred kpcs) that they will overlap at relatively large separations so that dynamical friction will lead to an ultimate merger.

More recently, 21cm velocities have been obtained for a sample of pairs extracted from the CfA catalog, including some at separations as large as 1Mpc (Chengalur et al. 1993). These more accurate redshift measurements are found to be crucial to the analysis of histograms of radial velocity differences for pairs. In fact the median velocity difference is $\sim 30$km/s for the Chengalur sample, comparable to the typical error in radial velocities in the CfA catalog. If galaxies were point masses on Kepler orbits the velocity difference could be this small for pairs at large separations, but should be considerably larger at small separations. A recent study (Chengalur et al. 1994) that focused on close pairs ($r_p < 75$kpc) found a similarly small median velocity difference, in contradiction with the expectations from Kepler orbits, though a larger sample is needed to analyze the significance of this result.

The goal of the study described in this paper was to conduct N-body simulations of galaxy pairs that could address the reasons for the various trends found in the data. Previous numerical treatments of galaxy mergers focused primarily on the phenomenology of the late stages of the merger process and the detailed features of the merger remnants (Toomre & Toomre 1972; White 1978, 1980; Gerhard 1981; Carlberg 1982; Villumsen 1982; Farouki, Shapiro, & Duncan 1983; Negroponte & White 1983; Barnes 1988, 1992). This study focuses on the orbital behavior of simulated galaxy pairs preceding merger with emphasis on the time spent at various projected separations and velocity differences, for particular choices of the extent of the dark matter halos and the initial orbital properties. Qualitatively, it seems possible that the small velocity differences that are observed for pairs at small separations (Charlton & Salpeter 1991; Chengalur et al. 1994) could be explained by the effect of dynamical friction. This effect will be most pronounced for the largest halos. On the other hand, large halos will have large masses, leading to larger infall velocities. Without doing quantitative simulations it is not clear which of these two effects will dominate, and thus whether dynamical friction is responsible for the observed small velocity differences of close pairs.

Since there is no observed increase in velocity with decreasing separation, the small velocity differences for wide ($\sim$ 1Mpc) pairs cannot be explained merely by bound Kepler orbits of point masses. Our models will also address how these values are produced. Without quantitative calculations of the effects of dynamical friction it is not clear if these small median velocity differences can result from the orbit of galaxies with large, overlapping, dark matter halos. The N-body simulation results are sampled at equal intervals over the various stages of evolution to produce simulated histograms of $r_p$ and $\Delta v$ that can be compared directly to the observations. Also, considering that timescales for orbits are comparable to a Hubble time, the alternative possibility that small velocity differences for wide pairs are due to a typical turnabout from Hubble flow at this distance will be assessed.

The following section of this paper will describe the initial conditions for the simulations. In section III the quantitative effects of dynamical friction for galaxy pairs on parabolic orbits will be studied as the dark matter halo radius is varied (for fixed distance of closest approach). The specific comparison of the simulations to the observed histograms of $r_p$ and $\Delta v$ (for various $r_p$ ranges) given in CS, and the median velocity differences given in Chengalur for wide and close pairs, follow in section IV. The final summary section will address the question of whether dynamical friction can produce the very small velocity differences for pairs in low-density regions at all separations, recently derived from 21cm data, (Chengalur et al. 1993, 1994) that were not previously apparent because of larger errors in the redshift catalogs, or because of less-stringent selection criteria (Rivolo & Yahil 1981; White et al. 1983; Charlton & Salpeter 1991; Schneider & Salpeter 1992). Specifically, we suggest what observations of galaxy pairs would be most likely to determine under what circumstances the various physical effects (turnabout, dynamical friction, Kepler orbits, dissipation due to gas) are dominant.

## 2. Initial Conditions for N-body Simulations

A series of simulations were performed in order to study the effects of dynamical friction on orbiting galaxy pairs. For the purpose of these simulations, each of the two galaxies is represented by 5000 equal mass particles in an isothermal density distribution. These particles approximate a spherical dark matter halo. Some number of central particles are designated as the luminous galaxy and their center of mass is followed for comparison with observed pairs. These dense central particles resist phase mixing despite the overlap of dark matter halos, and remain compact throughout the simulations, as was found in previous numerical work (White 1980; Barnes 1992). A three to one mass ratio was chosen for the two galaxies since there is typically a one magnitude difference between galaxies in the observed pair samples (CfA and SSRS) to which comparisons will be made. (This is appropriate if the mass to luminosity ratio is constant.) The individual particles in the more massive galaxy (galaxy A) have three times the mass of those in the less luminous galaxy (galaxy B) so that $M_{hA} = 3M_{hB}$.

Each particle in the halos is assigned a velocity with random direction and an initial speed equal to the root mean square velocity of an isothermal distribution $\sigma_h = \sqrt{M_h/R_h}$, where $R_h$ is the initial halo radius. The rms velocity of halo A is taken as the fundamental unit of velocity in these simulations. Since galaxy B is scaled down by a factor of 3 in mass, $\sigma_{hB} = \sigma_{hA}/3^{1/4}$ and $R_{hB} = R_{hA}/\sqrt{3}$.

Several papers have suggested that galaxy pairs which were initially on marginally-bound orbits may describe the current population of pairs most accurately (Toomre & Toomre 1972; White 1979; Villumsen 1982; Evrard & Yahil 1985; Schweizer 1987; Barnes 1992). Following this idea, we use initially parabolic orbits for the simulated pairs. This approach still does not incorporate "accurate" cosmological conditions, because galaxy pairs were never truly isolated. The simulations of cosmological clustering in an expanding medium, such as Evrard & Yahil (1985), which do tend to produce very eccentric orbits for binary galaxies, must be taken as complementary to the simulations of isolated galaxy pairs. Also, we use a spherical isothermal distribution of particles for each halo, whereas real galaxy halos may not be spherically symmetric, and may have significant infall zones which could affect the orbital evolution. Since we interpret the large-scale qualities of the orbits on timescales comparable to a Hubble time in this study, our results could be affected by the use of different initial conditions. However, this study is intended to yield general insights of the dependence of observed properties on global parameters of galaxy pairs. The effects of non-isolation on observed pairs, discussed in Chengalur et al. (1993), will also be addressed here.

The fundamental unit of length in the simulations is taken to be the parabolic distance of closest approach, $d_{clos}$. A series of four runs was conducted with a range of values of $R_{hA}$, listed in the second column of Table 1, in terms of $d_{clos}$. For each case the luminous radius of galaxy A, $R_{lumA}$, was taken as $0.2d_{clos}$, and the ratio of $R_{lumA}$ to the dark matter halo radius (the fractional number of "luminous particles") is listed in column 3. To justify this arbitrary choice, Cases 1 and 2 were repeated with a factor of three fewer particles designated as the luminous galaxy, and little difference was observed in the orbits of the centers of mass. The initial state of the halos is unrealistic in that there is a sharp boundary between the galaxy and empty space, but the initial separation was always set at $\geq 4R_{hA}$ to allow the halos to relax to a smoother, more realistic profile well before they begin to overlap.

The simulations were conducted using an hierarchical N-body tree code (Barnes & Hut 1986; Barnes 1995). In this program, the volume containing all particles was hierarchically subdivided into cells of progressively smaller linear size. The potential field at the position of particle $i$ for each timestep was approximated by the monopole and quadrupole moments of distant cells only if the distance to the cell exceeded the linear size of the cell by a factor of at least $f \sim 1.5$. (Our choice of the "tolerance parameter" $\theta \approx 1/f$ was 0.7; see Hernquist 1987 and Barnes 1995 for further details of this input parameter.) Thus more distant groups of particles were included in larger cells, reducing the number of calculations per particle so that the required CPU time increases only as NlnN. For nearby particles, the potential was calculated in the standard way:

$$\phi_{ij} = -\frac{Gm_j}{r_{ij}^2 + \epsilon^2}.$$

In the code, $G$ was set equal to one. The softening parameter, $\epsilon$, was set equal to the mean interparticle separation at the half-mass radius of galaxy B, since choosing a larger value would interfere with the accuracy of the tree code force approximations (Hernquist 1987), and a smaller value would increase fluctuations in the potential field (Hernquist & Barnes 1990). It is important to note that this softening parameter is comparable to the radius specified as the luminous galaxy, i.e. the interactions of the luminous galaxies are not being followed accurately. The calculation timestep for all particles in each case was set equal to $0.1 R_{lumA}/\sigma_{hA}$ at all times, and the energy was always conserved to better than 1% in the simulations (each of which ran in $\sim$ one week on a Sparc 10).

## 3. Simulation Results

The orbital tracks of the centers of mass of the particles representing the luminous galaxies are shown in Figure 1. Orbits are illustrated as the difference in the coordinates between the two galaxy centers. Dynamical friction causes a severe departure from the initial parabolic orbits as soon as the halos overlap. (A case was also run with $d_{clos}/R_{hA} = 4$. This case showed only a very small departure from the parabolic orbit due to the tidal transfer of bulk motion to the internal velocities of halo particles.) The galaxies trace out quasi-elliptical paths on the first one or two bounces after the first closest approach. A smaller eccentricity (more circular orbit) results from cases with less overlap (larger $d_{clos}/R_{hA}$). Larger dark matter halos lead to nearly radial bounces. There is little evidence in any of the cases for significant orbital circularization during subsequent bounces. This agrees with simulation results by Barnes (1992) which showed decreases in the orbital angular momentum of tightly bound particles after each passage which compensate the decrease in energy for the subsequent closer passages.

The evolution of the separation $R$ and the three-dimensional velocity difference $V$ of the luminous galaxies is shown in Fig. 2; note that the maximum values of separation correspond to the minima in the velocity difference and vice versa. The axes are labeled with distances in units of $d_{clos}$, velocities in units of $\sigma_{hA}$, and times in units of $t = R_{lumA}/\sigma_{hA} = 0.2 d_{clos}/\sigma_{hA}$. The successively decreasing humps in the graphs of $R$ as a function of time indicate the role of dynamical friction in converting bulk kinetic energy to thermal energy each time the galaxies overlap.

Without modeling the detailed structure of luminous galaxies it is difficult to decide exactly when a pair has merged, that is for practical purposes when it would be observed as a single galaxy in redshift catalogs. For our purposes, the merger time is indicated by two criteria: 1) the separation of the centers of mass of the luminous galaxies remained within about $2\epsilon$, 2) the rms velocity of the particles representing the luminous galaxies leveled to a constant value. These criteria were used to specify the maximum time for which pairs were included in the comparisons to observed pairs described in the next section.

The results can also be rescaled, individually for each case, such that the halo size (and mass) is constant and the distance of closest approach changes. This allows a comparison of the timescales of merger. Column 4 of Table 1 gives $d_{clos}$ in units of $R_{hA}$ (the inverse of column 2), and column 5 lists the time $\tau$ between first approach and merger (rescaled). The merger timescale, as expected, is an increasing function of $d_{clos}$, except for cases 3 and 4, wherein $d_{clos}$ is much smaller than the halo radii. In this regime, the collision is effectively head-on, and the merger timescale is constant with constant mass.

As discussed in the introduction, one of the important general facts to arise from observational pair studies is the small velocity differences that arise at small separations. We immediately see qualitatively from Figure 2 that the velocity differences at small separations are larger than the rms velocity of galaxy A, even when dynamical friction has its maximum effect during the closest approach of the galaxies.

The peak velocity at first encounter increases with increasing pair mass, but the increase is slow for the two most massive cases. The additional mass in case 4 (plot 2d) is placed at very large galactic radii, where it does not significantly deepen the potential well of either galaxy. This is equivalent to the similarity of the merger timescales of these two cases mentioned above. For a constant mass and changing $d_{clos}$, the smaller $d_{clos}$ will not move the pair into a significantly deeper portion of the potential well as long as $d_{clos} << R_{hA}$.

In examining Fig. 2 it is apparent that increasing the masses and sizes of galaxy halos does not always lead to a steady increase in the maximum separations during the various bounces. In fact the maximum separation is very similar for Cases 1 and 2 (Figs. 2a and b). This behavior is to be expected because of the two competing effects: 1) As the extent of the dark matter halo increases, the combined mass of the pair creates a larger potential well and the pair will be moving at a higher velocity at encounter and will achieve a greater subsequent maximum separation. 2) The deceleration of the galaxies near closest approach due to dynamical friction must be larger for larger galactic halos. To quantitatively examine this issue we give in Table 1 the maximum separation, $R_{max}$, achieved after the first approach (in units of $d_{clos}$). The role of dynamical friction must increase dramatically from case 1 to case 2, since case 2 has three times the total mass, and the two cases achieve the same $R_{max}$. Very roughly, the maximum separation can be quantified as $R_{max} = R_{hA} + 4d_{clos}$. This illustrates that if the distance of closest approach is small compared to the halo size, the increased acceleration on approach due to the increased mass will not be balanced by the increased effects of dynamical friction. It is only if $d_{clos}$ and $R_{hA}$ are comparable that the increasing halo overlap with increasing $R_{hA}$ will lead to a dominant effect of the increasing impact of dynamical friction on the subsequent orbit.

## 4. Comparison of N-body Results to Observed Pair Properties

In order to facilitate comparison of the simulation results to the observed distributions of pair projected separations $r_p$ and radial velocity differences $\Delta v$ several simplifying assumptions will be made. We will plot the histograms that are expected to arise if all of the observed pairs result from different stages of a particular type of orbit. In order that the orbit is equally populated at all stages, we must assume that enough time has elapsed that a pair could reach merger, and that the feed rate into the orbit is relatively constant over this period of time. In this section we will work in the context of these assumptions, but in the next section will discuss the problem in a more general context. Figure 3 shows scatter diagrams of $(r_p, \Delta v)$ that result from viewing each point $(R, V)$ in the simulations (Fig. 2) from a random angle. The data sets shown here are those from which the model histograms are generated.

Our simulation results can be scaled to correspond to the physical units of the observed pair samples. In the low density regions of the CfA and SSRS catalogs, the galaxy pairs have a median total luminosity of $8.9 \times 10^9 L_\odot$ (CS). Scaling with Milky Way values, the corresponding rotation velocity is $\sim 200$km/s for galaxy A (one velocity unit). For any of cases 1-4 a value of either $R_{hA}$ or $d_{clos}$ can be specified, and the physical time unit can then be determined.

### 4.1. Projected Separation Histograms

In CS it was determined that the majority of pairs in low density regions with $\Delta v < 150$km/s are bound, even at separations up to 1Mpc. The same velocity criterion is applied to select pairs from the model results for Cases 1-4 using two different choices for $d_{clos}$, and the resulting $r_p$ histograms are presented, along with the CS data, in Figure 4. For all cases, simulated pairs will only be included on the histogram if $\Delta v$ is less than .75 velocity units (or 150km/s). For $d_{clos} = 50$kpc, Cases 1 - 4 correspond to $R_{hA} = 50, 150, 500,$ and $1000$ kpc, and are shown in Fig. 4 a-d respectively. With a wider $d_{clos}$ of 200kpc, Cases 1-3 are rescaled to $R_{hA} = 200, 600,$ and $2000$ kpc (Figs. 4 e-g). Case 4 would scale to an unrealistically large halo size for this $d_{clos}$. The observed $r_p$ distribution (CS) is given for comparison in Fig. 4h. Although it is convenient to present the same binning scheme for all cases, it is clear that for fixed $R_{hA}$ the histograms could be scaled to any $d_{clos}$ just by compressing or expanding the horizontal axis. Sharp edges are apparent in the model histograms due to observations at the points of maximum separation where the pairs spend a large amount of time. The observed histogram is not expected to show such sharp features because it is likely to be an average of many values of galaxy mass and orbital parameters. Only the overall shape of the distribution should be compared. In addition, for real galaxy pairs at separations of more than 1Mpc, gravitational interaction with other nearby galaxies is more likely to be important. We shall thus focus on interpreting the histograms for $r_p < 1$Mpc.

The observed $r_p$ histogram (Fig. 4h) is approximately flat, with roughly the same number of pairs in each bin, despite the fact that the bins increase in area with increasing $r_p$. Thus the projected density of pairs at a given $r_p$ is proportional to $r_p^{-1}$ (the spatial density of pairs at separation $r$ is approximately proportional to $r^{-2}$). This represents an excess over the background level expected from a random distribution as illustrated with the solid line in Figure 4h. If we subtract this background level, the $r_p$ distribution declines slowly with increasing separation.

For the small distance of closest approach (50kpc) the lower mass, smaller halos (Figs. 4a-b) seem inconsistent with the data. This is due to the fact that the rebounds of the orbits do not go out to large separations and too much time is spent on bounces at small separations. Dark matter halo sizes larger than a few hundred kpcs (for this $d_{clos}$) are adequate matches to the data. The histograms of models with large $d_{clos}$ (Figs. 4e-g) all match the data reasonably well when the background level is subtracted from the observed distribution. Of all the cases shown in Fig. 4, the best matches to the observed data are provided by halo radii of 200kpc and 600kpc for the wider distance of closest approach, and halo radii of 1000kpc for the small $d_{clos}$ (Figs. 4d,e,f), although Figs. 4c and 4g cannot be excluded.

### 4.2. Radial Velocity Difference Histograms

Histograms of radial velocity difference in CS are given for various bins in $r_p$. In all of the bins, there is a peak at $\Delta v < 150$km/s. For relatively close pairs ($r_p < 150$kpc) the models are compared with the observed histogram in Fig. 5. (Again Cases 1-4 are scaled to $d_{clos} = 50$kpc to give Fig. 5 a-d, and Cases 1-3 are scaled to $d_{clos} = 200$kpc to obtain Figs. 5 e-g. The scaling only determines which pairs are included in the histogram, since the velocity unit is set equal to 200km/s in all cases.) It should be noted that typical errors in the CfA and SSRS catalogs are 25km/s, so that the true observed histogram is likely to be even narrower than the one in Fig. 5h. This question was addressed by the accurate 21cm measurements of Chengalur et al. (1993) that will be discussed in the following section. Here, the larger CS sample will be used to draw conclusions about the shape of the model histograms over the full range of $\Delta v$. Another consideration for interpretation is that the model histograms are by definition isolated, while this is not absolutely the case for the observations even for pairs in low density regions. The presence of non-isolated pairs leads to a larger fraction of $\Delta v$'s at larger values. This effect was analyzed by Chengalur et al. (1993) in subsamples with various isolation criteria. (There are expected to be only $\sim .5$ optical pairs per bin in $\Delta v$. This level is estimated by distributing at random the median number of galaxies in a large (5Mpc) region of the low density half of the CfA/SSRS catalogs.) We cannot rule out models that are lacking pairs at the larger $\Delta v$ since this discrepancy could always be resolved by adding to the models some triplet or quadruplet systems. However, we can draw a meaningful conclusion if a model produces too many pairs with large $\Delta v$.

All of the peaks at low velocity difference in the histograms in Fig. 5 are at least as narrow as that shown in the data (Fig. 5h). The disadvantage of the two cases with the largest halo radii (Figs. 5d,g) is that there are a relatively large number of pairs at high velocities (greater than 200km/s). This is due to the larger acceleration associated with the larger total pair mass. Successful models should, if anything, underproduce large $\Delta v$ compared to the observed data which are, at least to some extent, contaminated by galaxies in small groups. The cases with smaller halo radii (4a,b,e) can fit the data with the inclusion of some non-isolated pairs, and Figs. 5c,f also may match the observed distribution, although not too many additional galaxies in small groups with larger $\Delta v$ can be added to these cases.

The consideration of pairs at very small separations is suspect because of the possible dependence on the merger time. These models do not address the question of how long we can measure the luminous galaxies as separate entities or what the velocities of these galaxies should be once the luminous parts begin to overlap. This is particularly problematic in light of the fact that the softening parameter is comparable to the size of the luminous galaxies. Thus it may be more conclusive to consider the $\Delta v$ histograms for pairs with $150 < r_p < 400$kpc given in Figure 6. For this case the background level due to optical pairs is estimated as $\sim 3$ pairs per bin in $\Delta v$ (the solid line in Fig. 6h), still leaving a significant number of observed physical pairs with large velocity differences.

In the simulation histograms, several cases (particularly those with small $d_{clos}$ and small $R_{hA}$) show a very pronounced peak at small $\Delta v$. As discussed above, the disagreement of these with the observed histogram cannot generally be used to rule models out, because of the possible contribution at large $\Delta v$ due to members of small groups which the models have neglected. However, the discrepancy in these cases may be too large to explain. The largest halo size for $d_{clos} = 200$ ($R_{hA} = 2000$kpc, given in Fig. 6g) is inconsistent with the data because it has too many large $\Delta v$ pairs as compared to small $\Delta v$.

### 4.3. Median Radial Velocity Difference

Chengalur, Salpeter, and Terzian (1993) found a median $\Delta v$ which is smaller than that of previous studies that were more biased toward narrow pairs (Schneider and Salpeter 1992). Their pair sample was selected from the CfA survey and the SSRS, but has more accurate radial velocity measurements obtained at Arecibo and Parkes Observatories. For their pairs, with $r_p < 1$ Mpc and $\Delta v < 200$ km/s and isolated within .75Mpc and 300km/s, CST gave a normalized form of the observed $< \Delta v >$ for comparison to computer modeling. The radial velocity differences of the pairs were divided by the quantity $W_0 = (W_A^4 + W_B^4)^{1/4}$, where $W_A = 2\sigma_{hA}$ and $W_B = 2\sigma_{hB}$ are the velocity widths of each galaxy in the pair. The observed median $< \Delta v/W_0 >$ was 0.08 for the pairs with $\Delta v < 200$km/s. The median $< \Delta v/W_0 >$ is calculated, for a variety of model cases for which $W_0 = 2(4/3)^{1/4}\sigma_A = 2.14$ velocity units, and presented in column 3 of Table 2.

For all of the cases, the $<\Delta v/W_0>$ values are larger than the observed value of 0.08. This discrepancy was not apparent with the CfA and SSRS redshift measurements because of the larger errors, but the CST measurements show a considerably narrower peak at low $\Delta v$.

It is relevant to ask to what extent the smaller observed $<\Delta v/W_0>$ values (in the CST study as compared to earlier samples) were due to the additional wide pairs that were included. The CST scatter plot of $r_p$ vs. $\Delta v$ does not show a statistically significant correlation (Figure 7). However, because of the small number of observed pairs with $r_p < 400$kpc, the data are still consistent with the smaller value of $\Delta v$ for wide pairs than for narrow ones that is seen in the model scatter diagrams (Figure 3). Table 2 also gives median values for the models in the three ranges of $r_p$. For some models the $\Delta v/W_0$ median is as small as .08 for wide pairs. In particular $d_{clos} = 50$kpc and $R_{hA} = 500$kpc could still be completely consistent with the CST histograms if larger close pair samples yield a larger median $<\Delta v/W_0>$ at smaller $r_p$ than the entire wide pair sample.

## 5. Summary and Discussion

We have presented the results of a series of N-body simulations that address the consequences that dynamical friction has on the distribution of pairs of galaxies. Only the case of an isolated pair of galaxies was considered and the galaxies were taken to have a mass ratio of three to one in order to mimic the typical magnitude difference in observed samples. The ratio of the radius of the dark matter halo of the more massive galaxy $R_{hA}$ to the distance of closest approach $d_{clos}$ (expected from the initial parabolic orbit of the two galaxies) was the key parameter. These simulations allowed us to address the magnitude of the effect of dynamical friction for various size halos. In addition the simulation results were scaled to physical units for comparison to observed histograms of projected separations and radial velocity differences.

In the introduction we suggested that the small observed velocity differences for close pairs could be a consequence of deceleration due to dynamical friction. More generally the question was posed as to whether an increase in the size of the dark matter halo of a galaxy would lead to smaller velocity at closest approach due to the increased effects of dynamical friction or to a larger velocity due to the increased mass. In section III we examined the series of cases with $R_{hA}/d_{clos} = $ 1, 3, 10, and 20 (Cases 1 - 4). Table 1 shows the result that increasing the extent of the dark matter halo for a fixed distance of closest approach generally increases the maximum separation, $R_{max}$, achieved on the orbit after the first encounter of the two galaxies. However, when the halo extent was comparable to the distance of closest approach, we found that the increasing overlap of the dark matter halos compensated for the increase in mass. Thus the cases with $R_{hA}/d_{clos} = 1$ and 3 "rebound" to the same physical distance despite the increased halo mass in the latter case.

In Section IV, we scaled the simulation results for comparison to various specific observed samples, particularly considering the questions of whether a narrow or wide distance of closest approach is favored, and whether a large or small halo size is favored. The conclusions were:
1) The observed flat projected separation distribution is difficult to fit with small halo sizes for any $d_{clos}$ (in agreement with the qualitative claims of CS). For small $d_{clos}$ the histograms were too heavily peaked at small separations, while for large $d_{clos}$ a relatively large halo size is needed in order that the galaxies do not simply follow a Kepler orbit. 2) Radial velocity difference histograms were considered separately for close pairs ($r_p < 150$kpc in Fig. 5) and for intermediate pairs ($150 < r_p < 400$kpc in Fig. 6). Galaxies with very large radii are not favored by the $\Delta v$ histograms. If the galactic radius is greater than 1000kpc, there are likely to be too many pairs in the model at very high velocity differences. In general, the best fit cases for all the histograms have halo radii $\sim$ 200 - 600kpc and a large distance of closest approach, although the case of $R_{hA} = 500$kpc and $d_{clos} = 50$kpc cannot be ruled out. The radial extent of the dark matter halos of these best fit cases are in general agreement with the recent observational studies of satellite galaxies (White & Zaritsky 1992; Zaritsky et al. 1993). 3) All of the models produce a median $\Delta v$ (normalized to the pair luminosity) that is somewhat larger than that observed by CST.

We propose the following basic picture to explain these results. Because the time required to complete a merger on many of the orbits we have discussed is comparable to a Hubble time, it seems quite possible that some observed pairs are still at the beginning stages of orbits. So we suggest that once a galaxy pair begins on an orbit, the resulting distributions of separation and velocity difference could be as our models predict; however, extra time may be spent at large separations and small velocity differences when the pair is at the turn-about point, where gravity just balances the Hubble expansion. Some cases are already marginally consistent with the low observed median $\Delta v$'s, and the effect of turnabout can bring these cases into good agreement. Another possible reason for the low observed median $\Delta v$ is that real pairs exist preferentially near the point of maximum separation after first encounter, which violates our assumption that a typical pair orbit is equally populated at all stages. This would occur if the rate of infall of galaxies into pair orbits decreased with time during early epochs. Currently there would be fewer pairs in the initial approach stages than our models predict.

Because of the timescale problem, the inability to follow the luminous galaxy interaction, and uncertainty about when merger actually occurs, it is difficult to reach definite conclusions regarding the halo size and the orbital parameters. However, this study has yielded some general insights that should be emphasized and can be used to guide future observations.

Turn-about can be partly responsible for the small $\Delta v$ pairs observed by CST, but this will apply only to wide pairs. The CST pairs study has only a small number of isolated pairs with $r_p < 400$kpc. If turnabout is not a significant effect at intermediate separations, the models should predict the correct $\Delta v$ in the $150 < r_p < 400$kpc regime. If the observed median $\Delta v$ is as small in this regime as at larger or smaller $r_p$, then dynamical friction cannot be responsible for the observed small $\Delta v$'s and some additional effect must come into play. Accurate radial velocity measurements for a larger sample of pairs with $150 < r_p < 400$ kpc would allow us to resolve this issue.

The recent study of narrow pairs (Chengalur et al. 1994) focuses on $r_p < 75$kpc; this study yields a very small median $\Delta v$, also $\sim 30$km/s, for the six sample pairs. The values in Table 2 for pairs with projected separations less than 150kpc are larger than this value by approximately a factor of two. This discrepancy could be a consequence of the fact that our simulations do not include the effects of dissipation (Negroponte & White 1983; Kormendy 1989; Noguchi 1991). Our simulations have shown that stellar dynamical friction should not yield a median $\Delta v$ as small as the observed value for close pairs. However, if the galaxies are interacting, gas dissipation effects may decrease the $\Delta v$ measured from neutral hydrogen radial velocities, as is the case for pairs in the Chengalur et al. (1994) study, without producing a net drag on the stellar components of the galaxies, which only respond to gravity. All the pairs in that study do show signs of interaction, and none have $\Delta v > 60 km/s$. Further observations of narrow pairs are essential to address the question of how often a very large $\Delta v$ can arise from a pair that shows signs of interaction, and possibly to find a discrepancy between the systemic velocities of the gaseous and stellar components. Despite the fact that Kepler infall and the usual random motions of galaxies predict $\Delta v$ comparable to the rotation velocity of a galaxy, the data thus far seem to show that the more precise the measurement, the smaller the velocity difference.

We are grateful to Jayaram Chengalur and Edwin Salpeter for their insightful suggestions, and also to Joshua Barnes for generously providing the N-body tree code, and for his helpful comments and suggestions for revision.

Table 1: Initial conditions and properties
of the simulations.

|  | $R_{hA}(d_{clos})$ | $R_{lumA}/R_{hA}$ | $d_{clos}(R_{hA})$ | $\tau(\frac{R_{hA}}{\sigma_{hA}})$ | $R_{max}(d_{clos})$ |
|---|---|---|---|---|---|
| Case 1 | 1  | 0.2    | 1    | 50  | 6  |
| Case 2 | 3  | 0.0667 | 0.3  | 12  | 6  |
| Case 3 | 10 | 0.02   | 0.1  | 7.5 | 14 |
| Case 4 | 20 | 0.01   | 0.05 | 7.5 | 27 |

Table 2: The normalized median of $\Delta v$ for all cases,
and for various bins of $r_p$.

| | $d_{clos}$ (kpc) | $<\Delta v/W_o>$ | | | |
|---|---|---|---|---|---|
| $r_p$ range (kpc) | | 0 - 1000 | 0 - 150 | 150 - 400 | 400 - 1000 |
| Case 1 | 50  | 0.10 | 0.13 | 0.09 | 0.10 |
| Case 2 | 50  | 0.14 | 0.13 | 0.10 | 0.17 |
| Case 3 | 50  | 0.11 | 0.15 | 0.13 | 0.08 |
| Case 4 | 50  | 0.13 | 0.17 | 0.11 | 0.12 |
| Case 1 | 200 | 0.11 | 0.10 | 0.17 | 0.10 |
| Case 2 | 200 | 0.12 | 0.11 | 0.15 | 0.12 |
| Case 3 | 200 | 0.14 | 0.17 | 0.11 | 0.15 |

## Figure Captions

Figure 1: Plots of the orbital tracks for the four cases. The solid curves trace the coordinate differences of galaxy B around galaxy A, and the dashed curves show the parabolic trajectories of point masses corresponding to the masses of each case. The axes are in units of $d_{clos}$.

Figure 2: The spatial separation of the centers of mass of the luminous galaxies, and the three-dimensional velocity difference, versus time. The curves are truncated at the estimated time of merger. The units of R and V are the fundamental units $d_{clos}$ and $\sigma_{hA}$, respectively, as discussed in section II.

Figure 3: Scatter diagrams of $(r_p, \Delta v)$ generated from the models. The axes are in units of $d_{clos}$ and $\sigma_{hA}$. The density of points is reduced by a factor of three from the original data sets.

Figure 4: Projected separation histograms for the subset of pairs with $\Delta v$ less than 150 km/s. Figs. 3a-g show the model distributions for the four cases, with two different choices of $d_{clos}$. The width of each bin is 0.1 Mpc. The left axes show the frequency of occurrence of values normalized to the total number of points in the histogram. For convenience, the scaled value of $R_{hA}$ is shown in each histogram. In figures 4d and 4g, the values are 1 Mpc and 2 Mpc, respectively. The values of $R_{max}$ are also shown where they appear in the limits of the histograms. Figure 3h is the observed sample; the right axis shows the total number of pairs in a bin. The solid line represents the distribution of pairs expected from a purely random distribution (CS).

Figure 5: Radial velocity difference histograms for the subset of pairs with $r_p$ less than 150 kpc. The width of each bin is $\sim$ 33 km/s. The left axes show the frequency of occurrence of values normalized to the total number of points in the histogram. Figure 4h is the observed sample; the right axis shows the total number of pairs in a bin. The solid line represents the distribution of pairs expected from a purely random distribution (CS).

Figure 6: Radial velocity difference histograms for the subset of pairs with 150 kpc $< R_p <$ 400 kpc. The width of the bins is $\sim$ 33 km/s. The left axes show the frequency of occurrence of values normalized to the total number of points in the histogram. Figure 5h is the observed sample; the right axis shows the total number of pairs in a bin. The solid line represents the distribution of pairs expected from a purely random distribution (CS).

Figure 7: Scatter diagram of observed isolated galaxy pairs from Chengalur, Salpeter, & Terzian (1993).

Figure 1

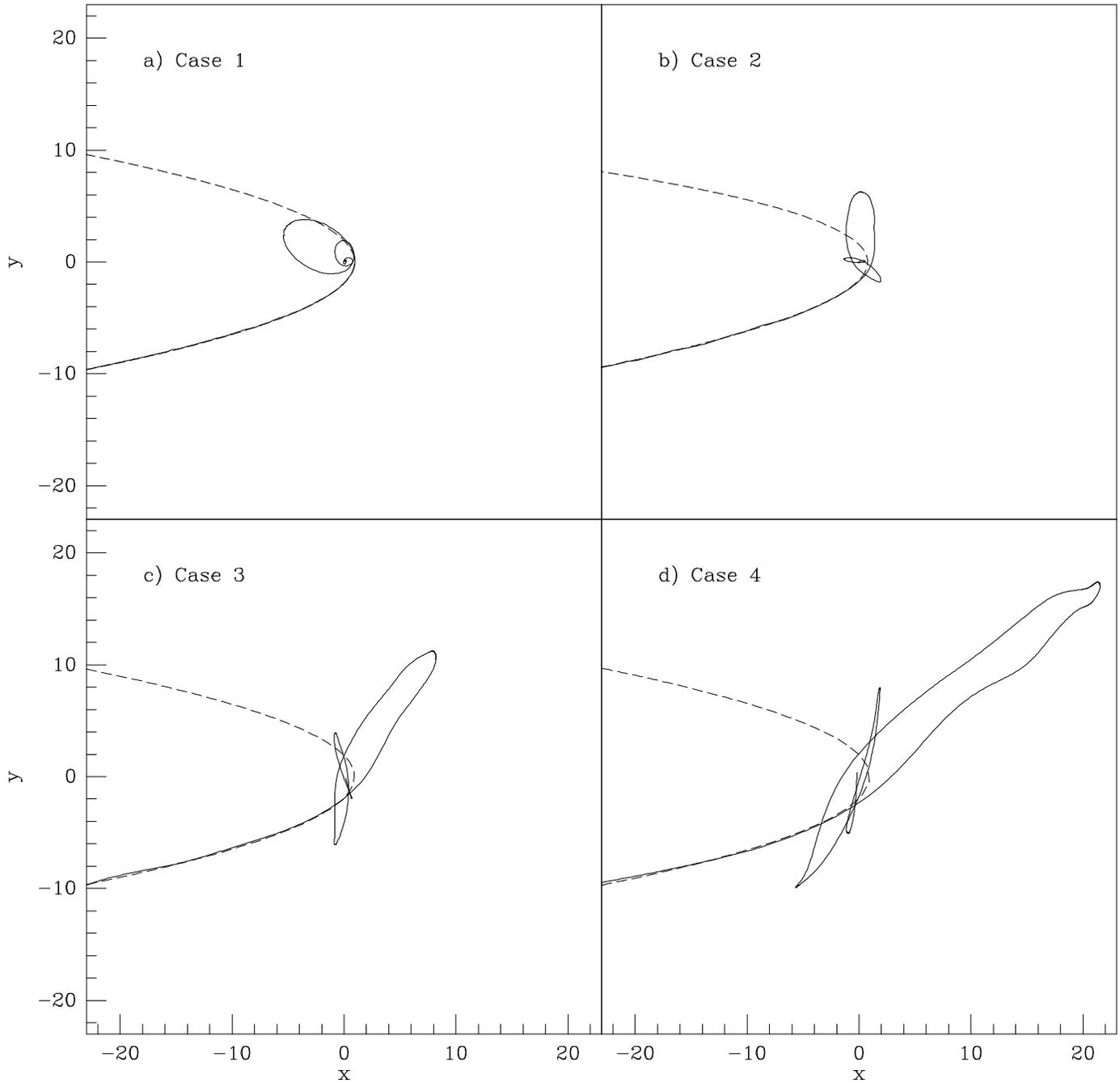

Figure 4

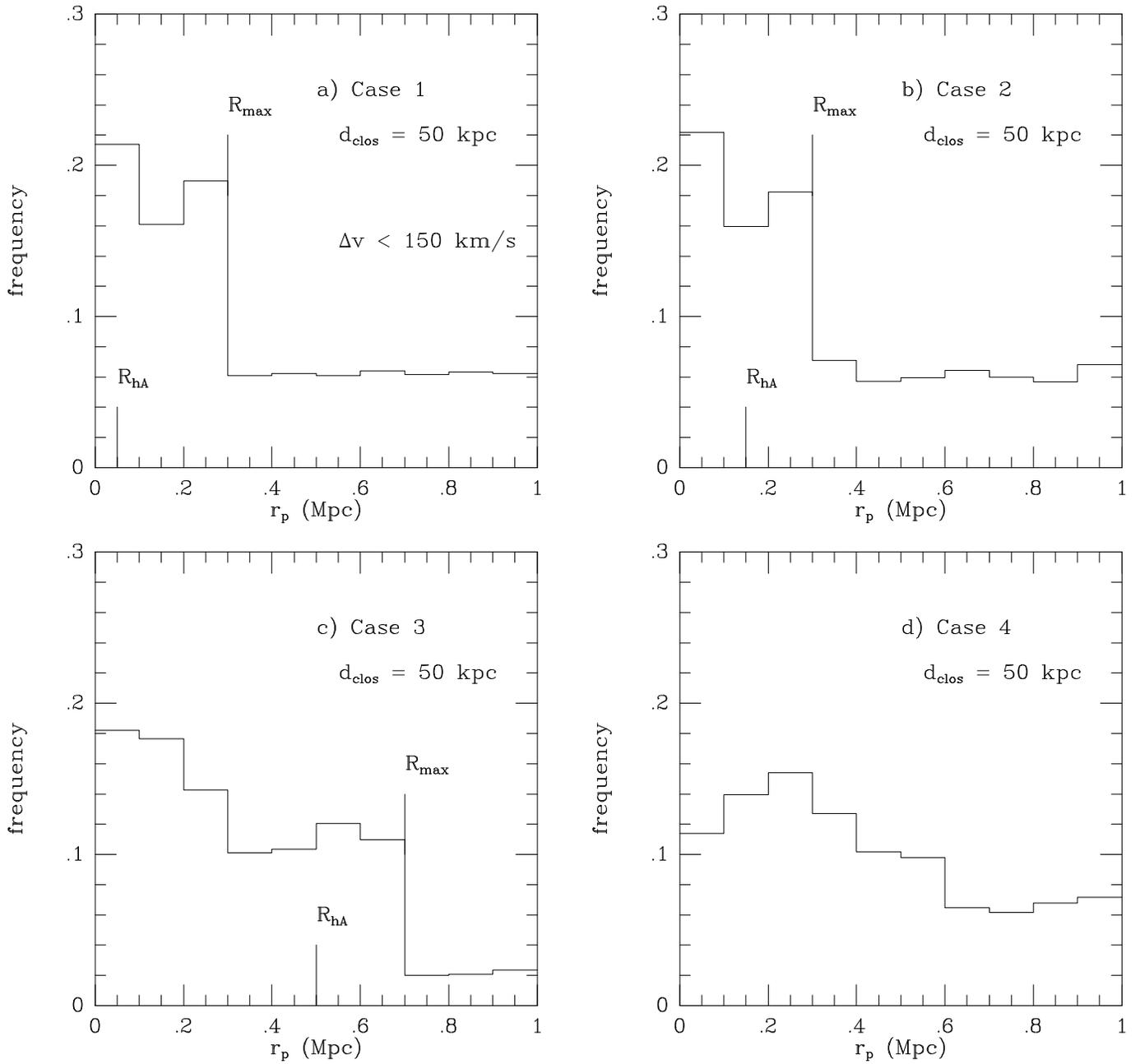

Figure 5

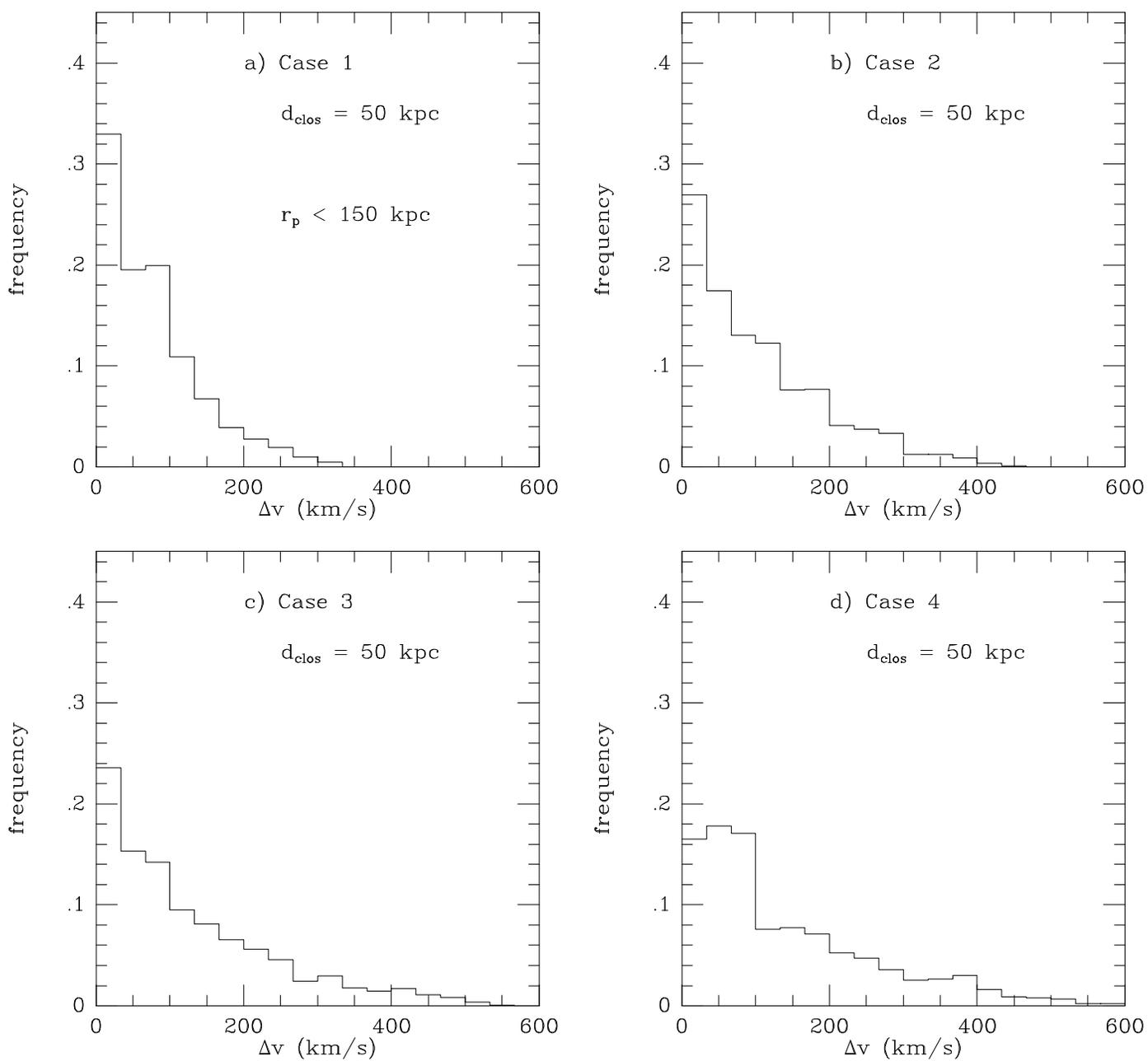

Figure 6

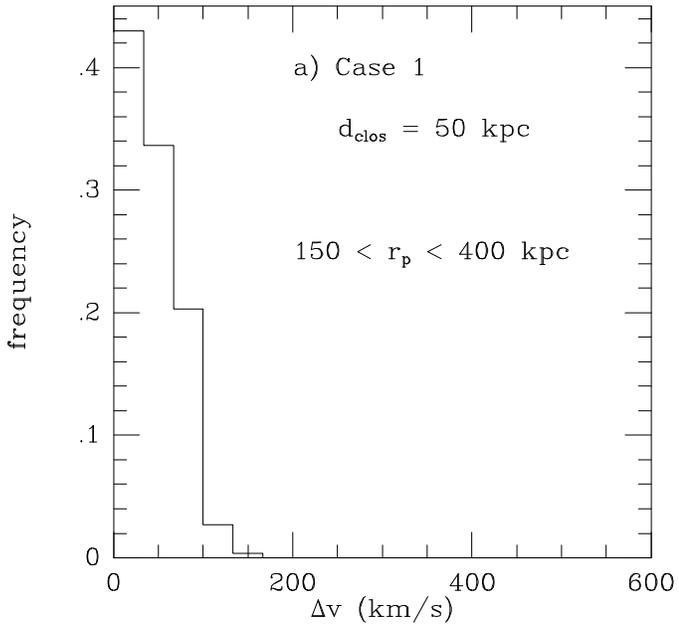
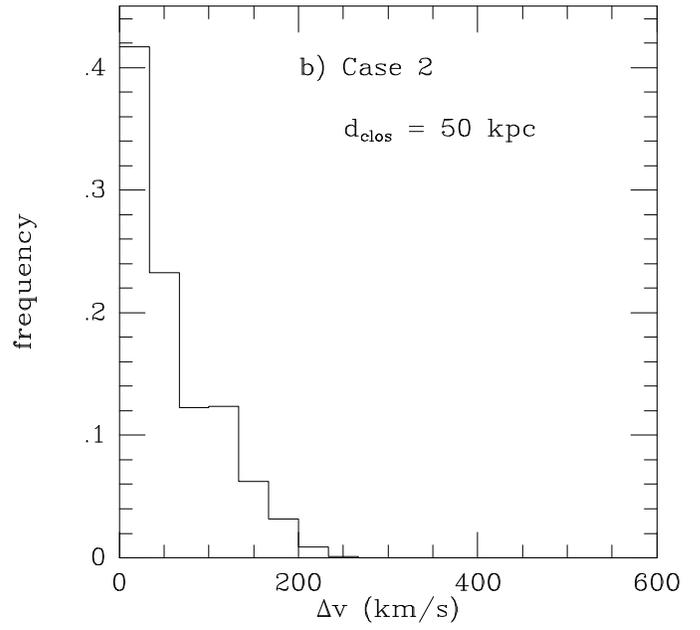
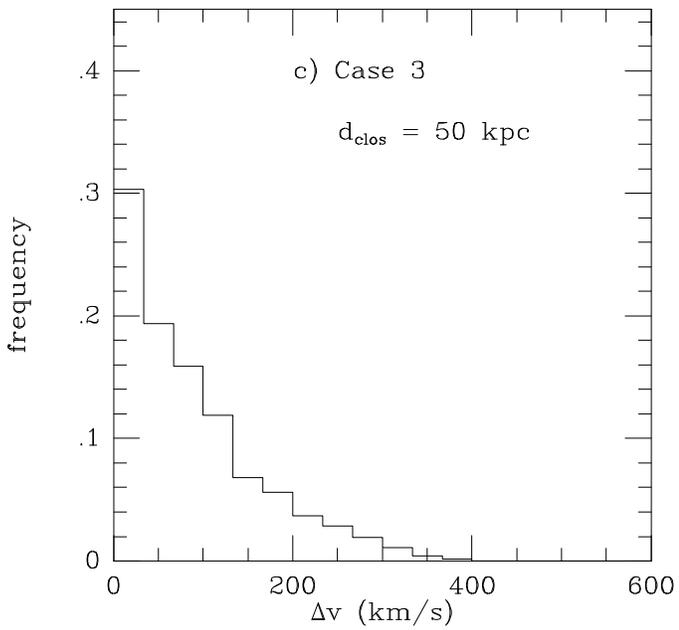
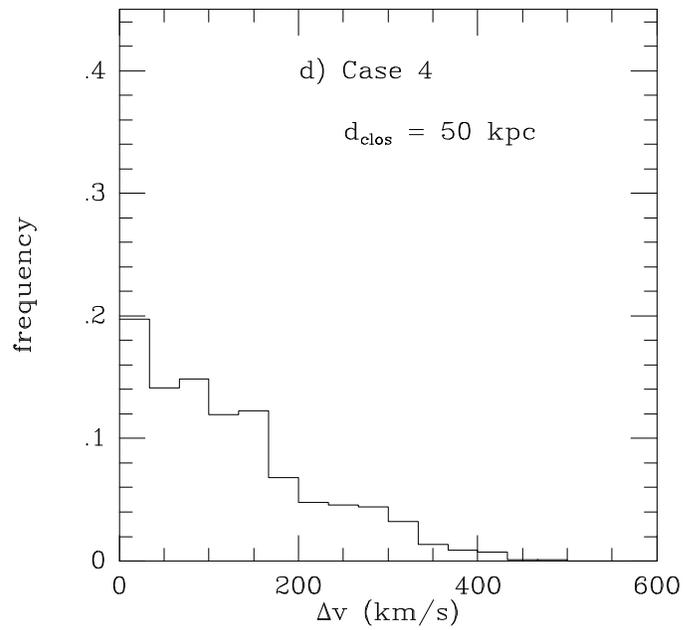

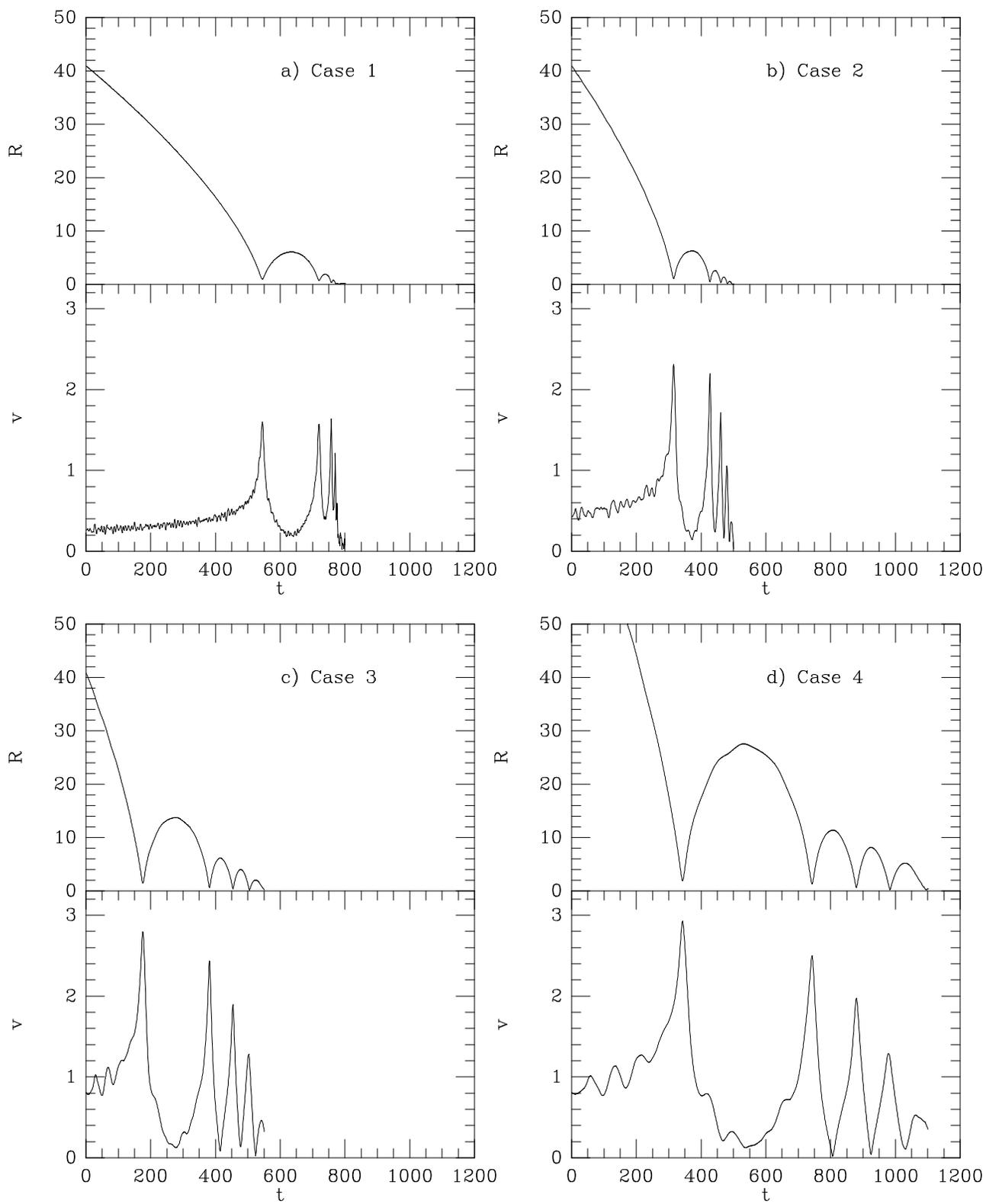

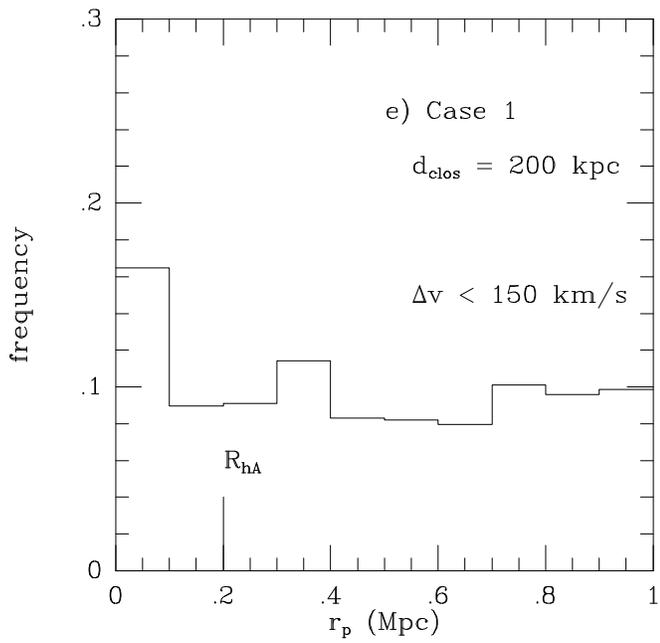
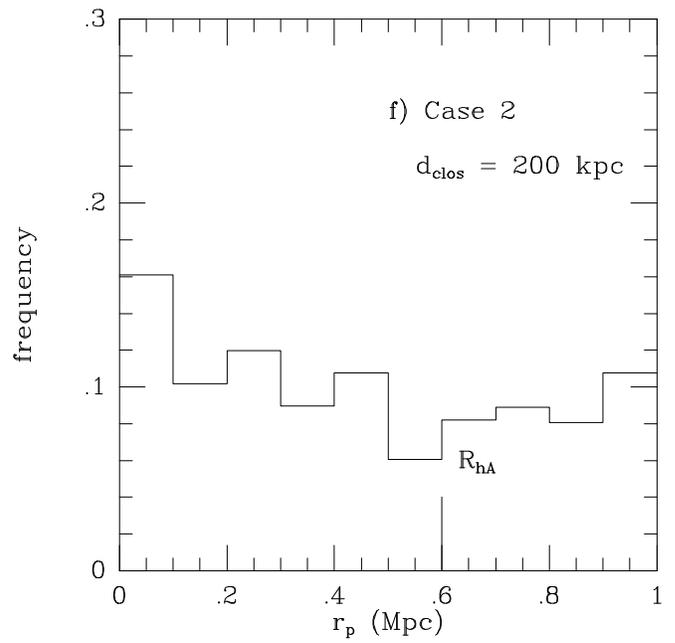
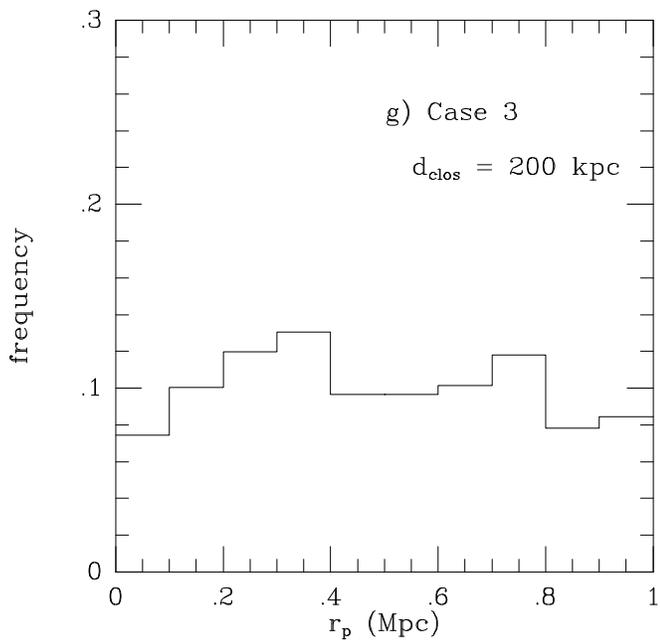
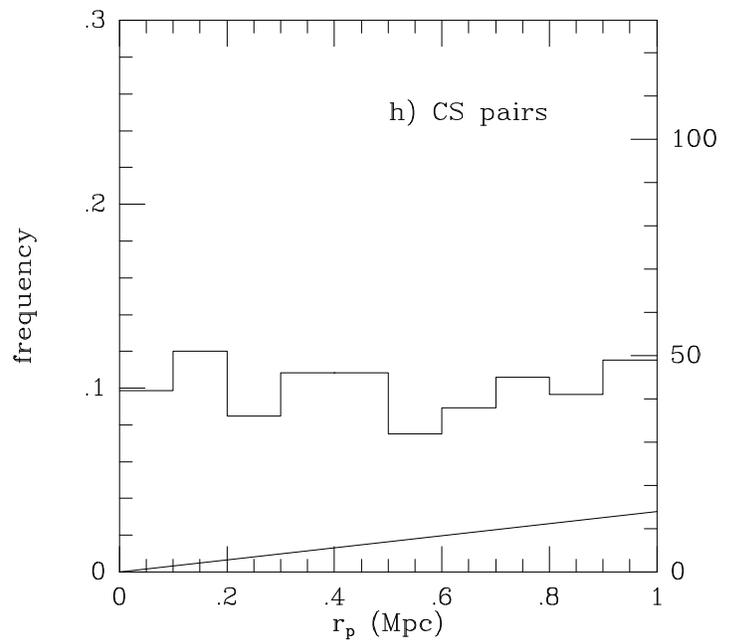

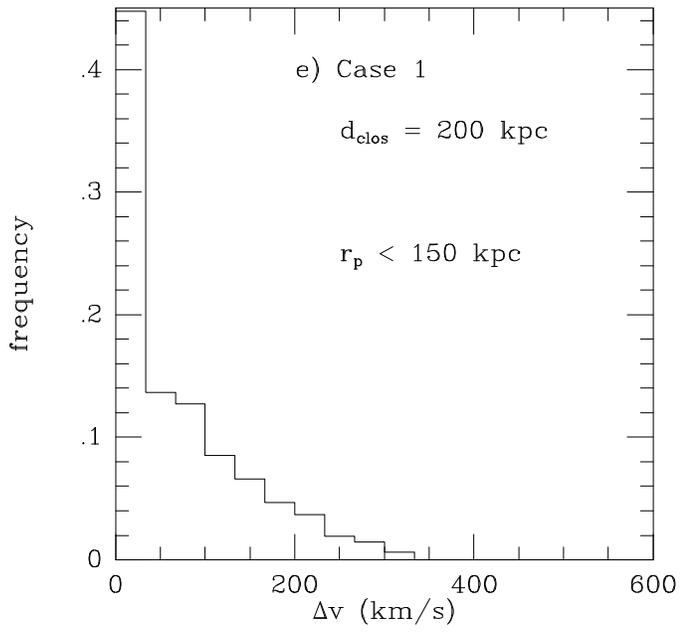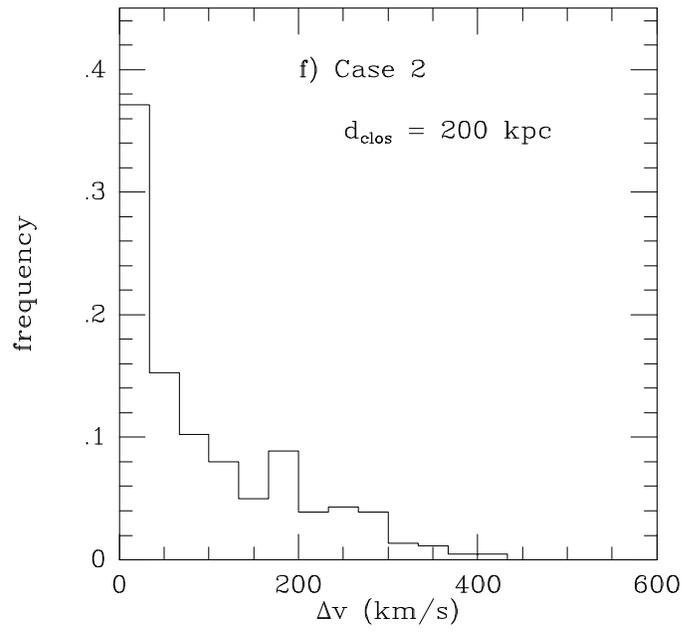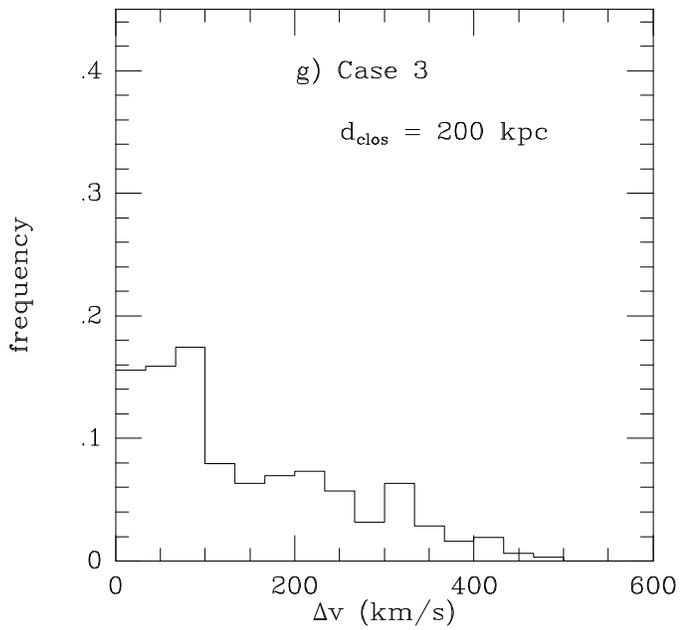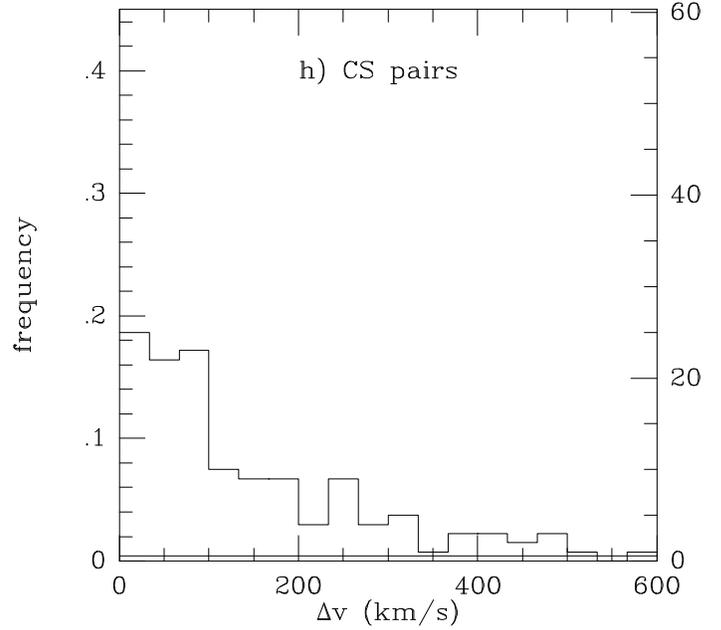

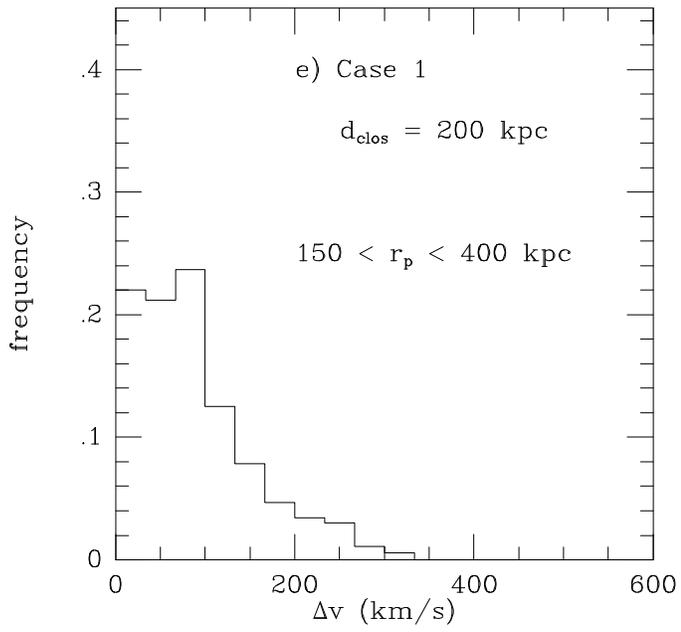
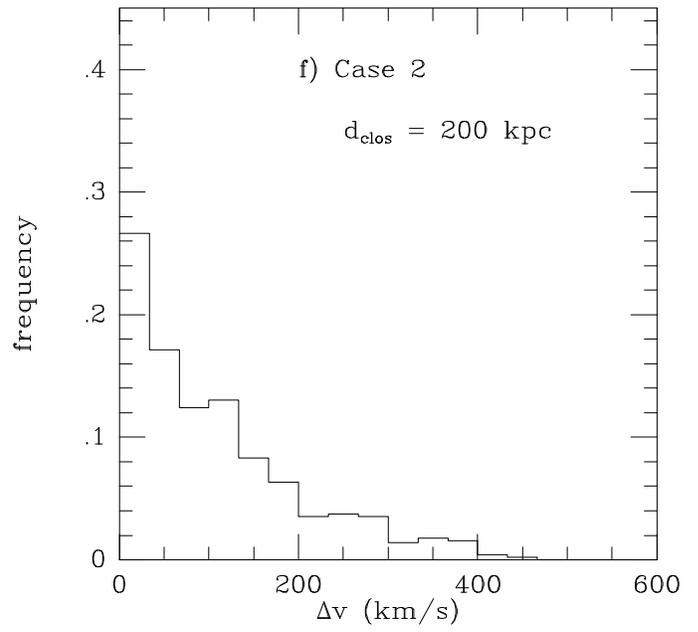
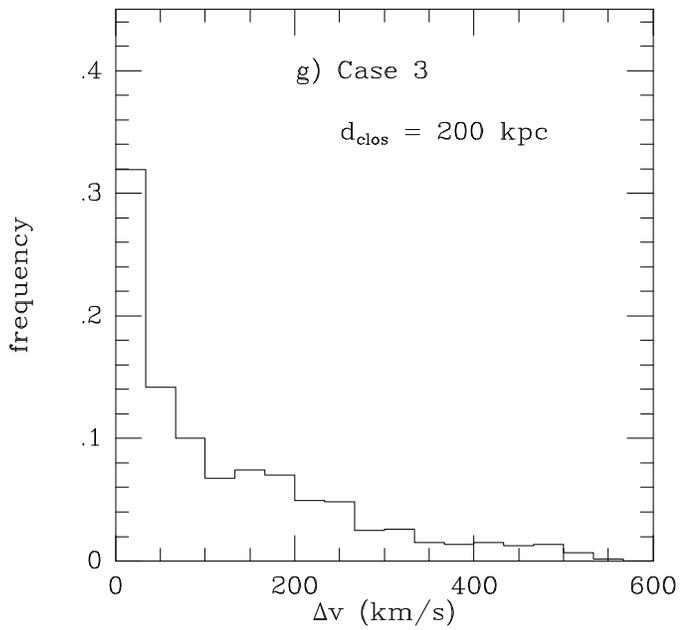
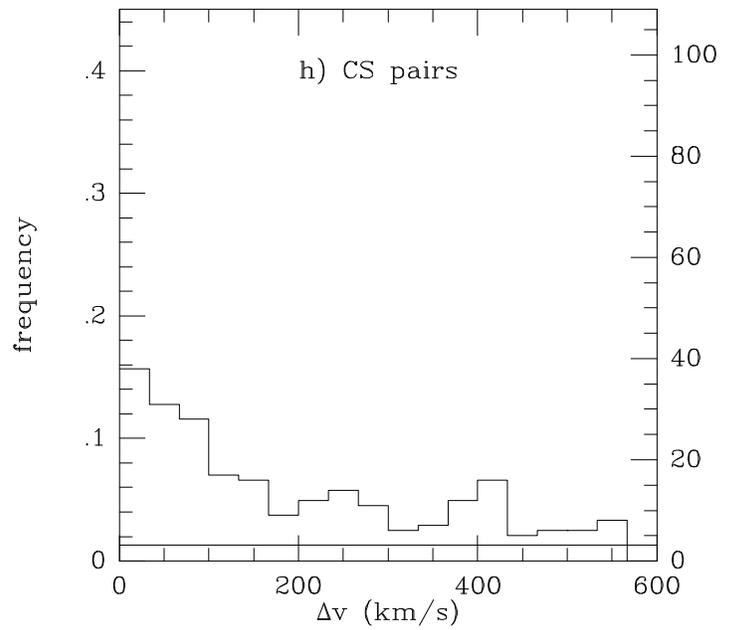

Figure 3

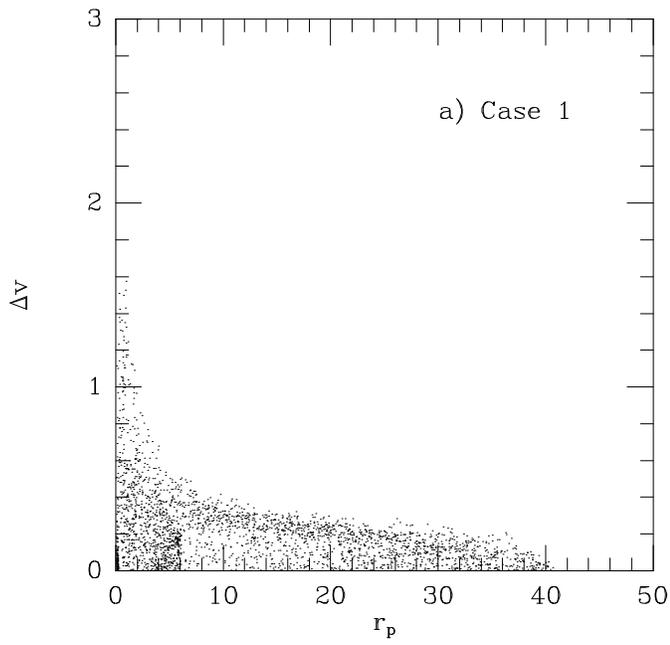
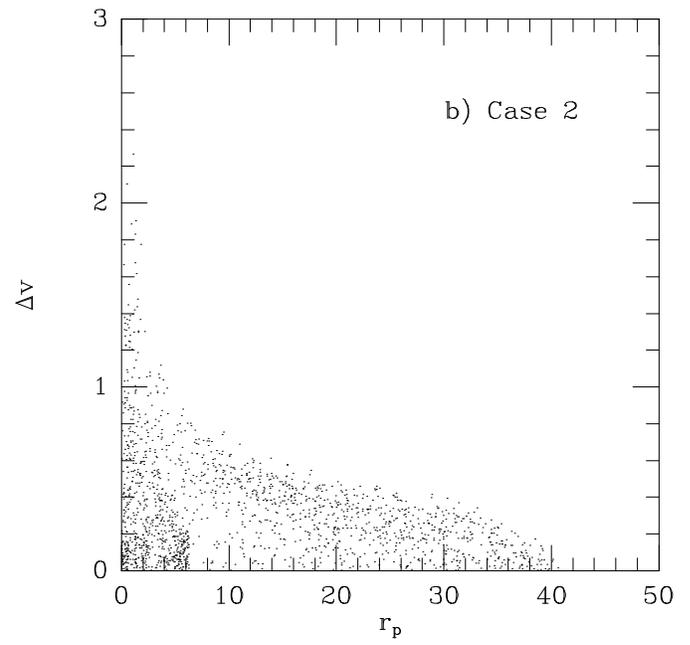
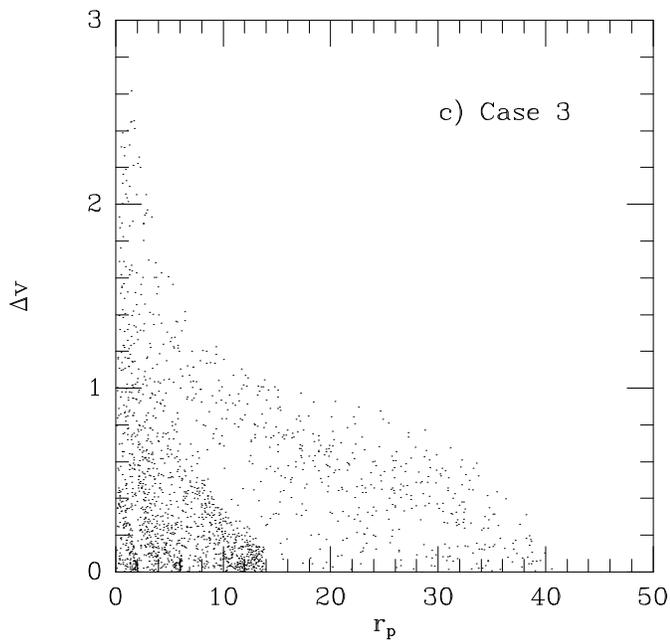
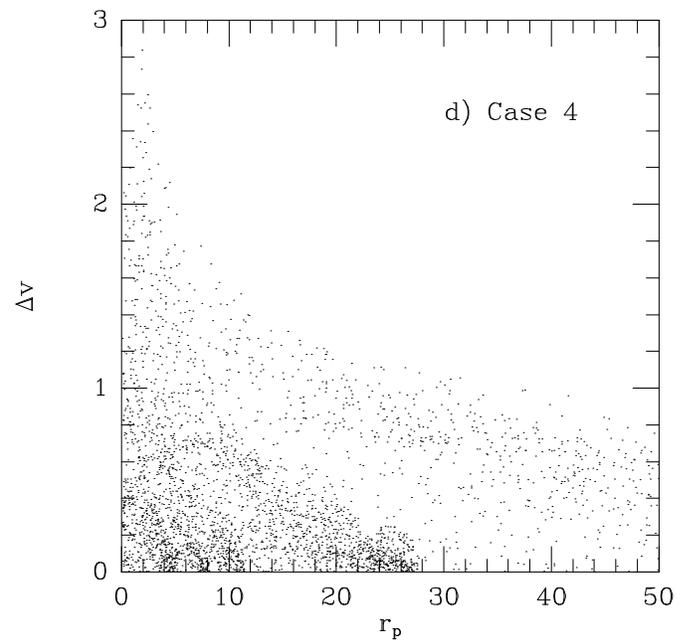

Figure 7

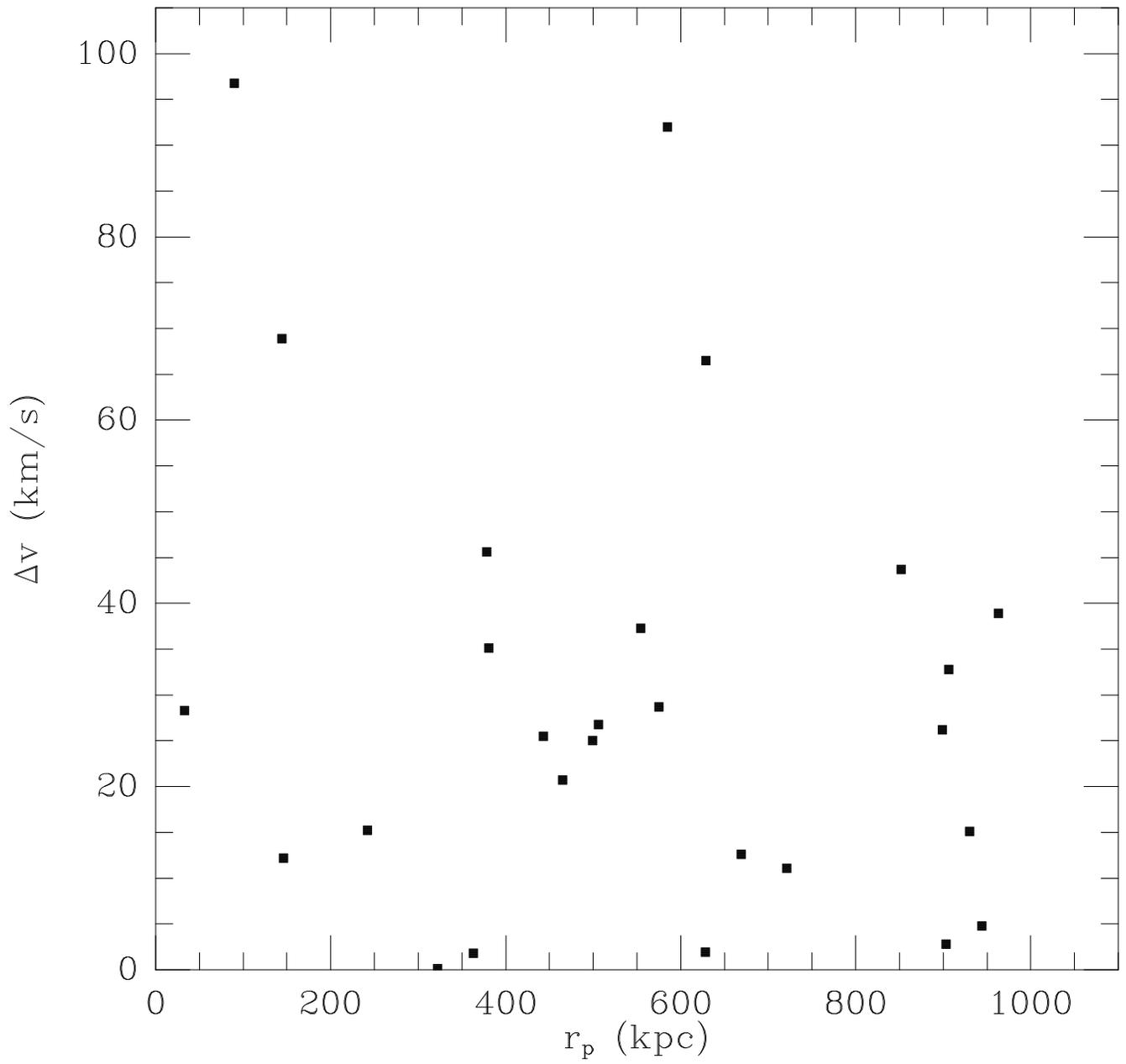